\documentclass[10pt]{iopart}
\usepackage{iopams}
\pdfoutput=1
\usepackage{graphicx}
\usepackage{cite}
\renewcommand{\vec}[1]{\mbox{\protect\boldmath$#1$}}
\begin{document}
\title[Off-equatorial orbits in strong gravitational fields near compact objects]{Off-equatorial orbits in strong gravitational fields near compact objects}

\author{Ji\v{r}\'{i} Kov\'{a}\v{r}$^1$, Zden\v{e}k Stuchl\'{i}k$^1$ and Vladim\'{i}r Karas$^2$}
\address{
$^1$Institute of Physics, Faculty of Philosophy and Science, Silesian University in Opava, Bezru\v{c}ovo n\'{a}m. 13, CZ-746 01 Opava, Czech Republic\\
$^2$Astronomical Institute, Academy of Sciences, Bo\v{c}n\'{i} II, CZ-141\,31\,Prague, Czech Republic
}
\ead{Jiri.Kovar@fpf.slu.cz}

\begin{abstract}
Near a black hole or an ultracompact star, motion of particles is governed by strong gravitational field. Electrically charged particles feel also electromagnetic force arising due to currents inside the star or plasma circling around. We study a possibility that the interplay between gravitational and electromagnetic action may allow for stable, energetically bound off-equatorial motion of charged particles. This would represent well-known generalized St\"{o}rmer's  \textquoteleft halo orbits\textquoteright, which have been discussed in connection with the motion of dust grains in planetary magnetospheres. We demonstrate that such orbits exist and can be astrophysically relevant when a compact star or a black hole is endowed with a dipole-type magnetic field. 
In the case of Kerr-Newman solution, numerical analysis shows that the mutually connected gravitational and electromagnetic fields do not allow existence of stable halo orbits above the outer horizon of black holes. Such orbits are either hidden under the inner black-hole horizon, or they require the presence of a naked singularity.  
\end{abstract}

\pacs{04.20.Jb, 04.70.Bw, 04.25.Nx, 04.40.Nr}

\maketitle

\section{\label{sec:Intro}Introduction}
Investigation of charged particle motion in strong gravitational and electromagnetic fields ranks among the elementary exercises in theoretical physics and astrophysics related to black holes or compact stars \cite{Pra:1980,Pra-Sen:1994,Sen:1997,Kar-Vok:1991,Vok-Kar:1991a,Fel-Sor:2003}. 
It helps us to reveal the structure of force fields acting on the particles. Numerical integration  of the motion equations is very helpful, but often fails to give full understanding. On the other hand, analytical techniques are often restricted, for practical reasons, to the motion along symmetry axis \cite{Bic-Stu-Bal:1989}, and to the equatorial \cite{Bal-Bic-Stu:1989} or spherical motion \cite{Joh-Ruf:1974}, or to the special case of free-infalling particles \cite{Stu-Bic-Bal:1999}. The general off-equatorial motion in the field of rotating black holes or compact stars still remains an interesting open problem deserving attention. In this paper, we tackle this problem focusing on the case of halo orbits, i.e., off-equatorial  energetically bound orbits that are stable with respect to small perturbations, but they never cross equatorial plane.   

As a motivation of our work, we mention one of the milestones of the classical space physics in this branch, i.e., St\"{o}rmer's analysis of charged particles motion in a purely dipole magnetic field (classical St\"{o}rmer problem) \cite{Sto:1955}, providing us with the basic physical description of radiation belts surrounding a magnetized planet. Radiation belts are known to be composed of individual ions and electrons whose motion is governed by magnetic forces only. In order to describe the dynamics of charged dust grains in planetary magnetospheres, when there are much smaller charge to mass ratios, one can employ the \textquoteleft generalized St\"{o}rmer problem\textquoteright, where both the planetary gravity and co-rotational electric field play a role \cite{Dull-Hor-How:2002}. Such studies point out the existence of the dust grains halo orbits near, e.g., Saturn \cite{How-Hor-Ste:1999}.    

Could such halo orbits survive in strong gravitational fields near compact objects with additional magnetic or electric fields? The situation is not obvious in case of exact solutions of Einstein-Maxwell equations, where the gravitational and electromagnetic fields are mutually interconnected. Surprisingly enough, this problem has not yet been fully answered in the case of well-known Kerr-Newman black hole.
In fact, we can anticipate such orbits to occur near the black hole horizon, where gravity is strong. Their exact location should depend on the type of particles we consider, namely, on the charge to mass ratio. Halo orbits could have outstanding astrophysical consequences for radiation of accreting compact objects. This is because the halo orbits form two lobes of stable motion - one above and the other one below equatorial plane - through which the accretion disc radiation has to pass on its way towards a distant observer. Therefore, the halo lobes act as a part of radiation reprocessing corona.

We investigate the problem with three qualitatively different kinds of compact objects. These are magnetized compact star, studied in terms of pseudo-Newtonian model (section~\ref{sec:CompactStar}), Schwarzschild black hole with a current loop (section~\ref{sec:Schwarzschild}), and Kerr-Newman black hole or naked singularity (section~\ref{sec:Kerr-Newman}). The first case is most close to the original St\"{o}rmer's approach, except that we consider strong gravity. The second case may serve as a toy model of the spacetime associated with a magnetized disc about a Schwarzschild black hole. In the last case, the solution was partly found in \cite{Fel:1979,Cal-Fel-Fab-Tur:1982}, where the authors conclude that the halo orbits do exist in Kerr-Newman spacetimes. Nevertheless, because of many input parameters characterizing the central object and the particle motion itself, the detailed discussion of the orbit existence was not presented there. We can remark that the rotating black hole immersed in a magnetic field will establish a stationary system with a non-zero net charge \cite{Wal:1974}. In astrophysically realistic magnetic fields, this charge is probably very small. However, we adopted general approach to the problem, which allows us to explore the whole parameter space in a systematic way. Thus, we considered situations in which the charge is small, but also those with large charge for completeness of our discussion. 

The paper \cite{Stu-Hle:1998}, dealing with a similar problem, i.e., with the stability of charged spherical shells in Kerr-Newman spacetimes, suggests that the existence of stable halo orbits is restricted to the naked-singularity spacetimes only. The authors of that paper employed the traditional four-dimensional covariant form of the motion equations. 
Our complete discussion of the Kerr-Newman spacetimes is based on inertial forces analysis \cite{Abr-Nur-Wex:1995,Agu-etal:1996,Stu-Hle-Jur:2000,Kov-Stu:2007}, which, in combination with the effective potential, gives a semi-analytic condition for the halo orbits existence.  

\section{\label{sec:CompactStar}Compact star with dipole magnetic field}
We consider a rotating object of mass $M$, radius $R$ and spin $\vec{\Omega}$ oriented in positive $z$-direction ($z+$). The object is a source of central gravitational field (potential $\psi$) described in terms of Gaussian units and Cartesian coordinates $(x,y,z)$. Further the object is endowed with rigidly co-rotating aligned dipole magnetic field $\vec{B}=\nabla\times\vec{A}$.\footnote{Electric and magnetic fields are related by the force-free relation $\vec{E}+(1/c)\vec{V}\times\vec{B}=0$, where $\vec{V}=\vec{\Omega}\times\vec{r}$.} Magnitude of the magnetic field measured at the object equator is $B_0$, and the induced co-rotating electric field has intensity $\vec{E}=-\frac{1}{c}(\vec{\Omega}\times \vec{r})\times \vec{B}$, where $\vec{r}=(x,y,z)$. 

Similar set up of the problem was employed in \cite{Dull-Hor-How:2002}, but here, we describe gravitational field of the central object by the Paczynski-Wiita potential \cite{Pac-Wii:1980}
\begin{eqnarray}
\psi=\frac{-GM}{r-r_{\rm S}}, 
\end{eqnarray}
where $r_{\rm S}=2GM/c^2$ is the Schwarzschild radius. Further, we use the common form of the dipole vector potential 
\begin{eqnarray}
\vec{A}=\frac{\mathcal{M}}{r^3}(-y,x,0),
\end{eqnarray}
where $\mathcal{M}=B_0R^3$ is magnitude of the $z+$ oriented dipole magnetic moment. 

\begin{figure}[t!]
\begin{center}
\includegraphics[width=0.75\hsize]{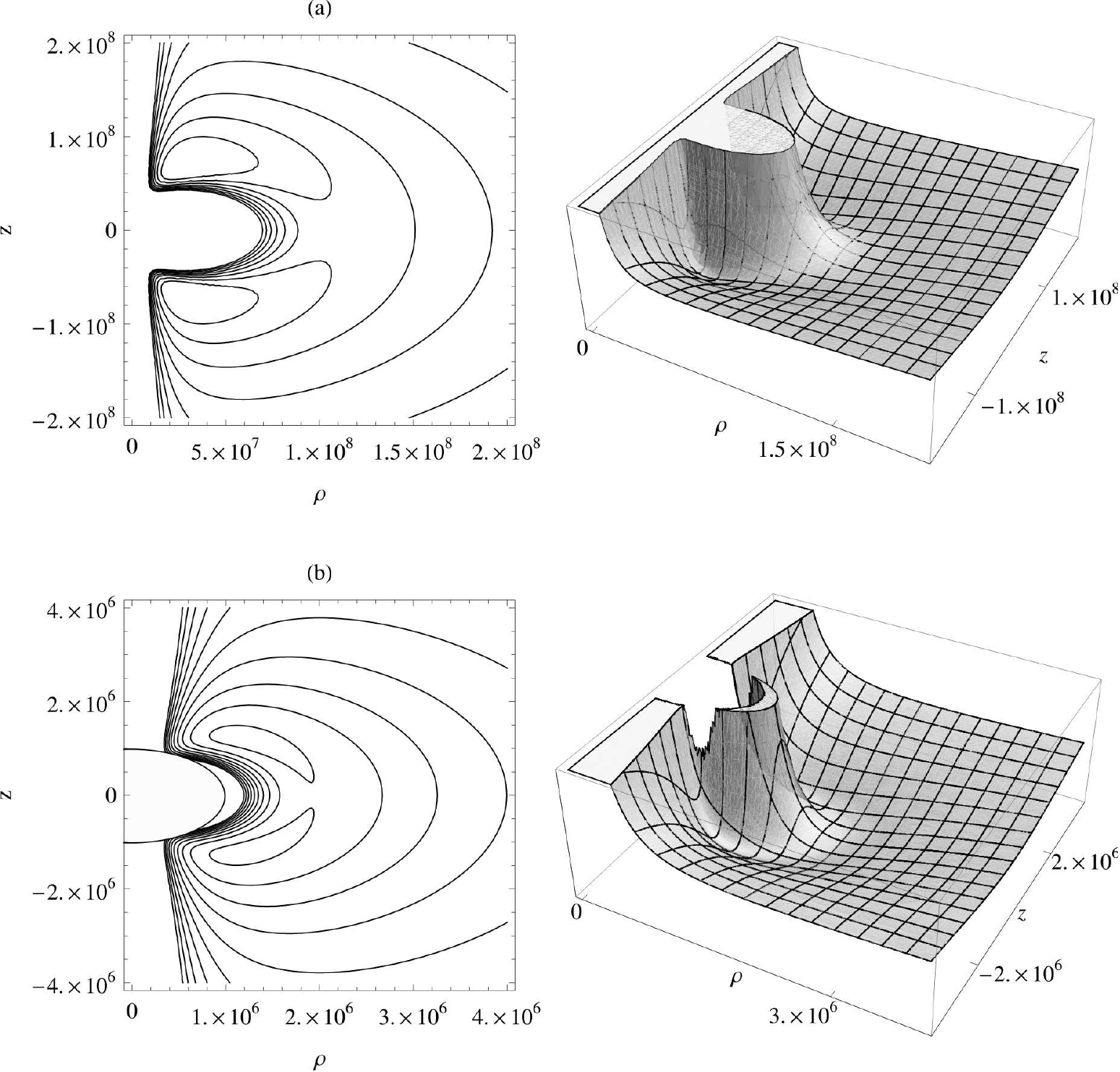}
\end{center}
\caption{\label{Fig:1} Effective potential $U_{\rm eff}$ and its contours for motion of particle with (in Gaussian units) {\bf (a)} $m=1.67\times10^{-24}$, $q=4.85\times10^{-10}$, $\bar{p}_{\phi}=-3\times10^{-8}$ (counter-rotating proton), or equivalently $q=-4.85\times10^{-10}$, $\bar{p}_{\phi}=3\times10^{-8}$ (co-rotating hydrogen anion); {\bf (b)} $m=4.2\times10^{-12}$, $q=-3.3\times10^{-5}$, $\bar{p}_{\phi}=2.5\times10^{4}$ (co-rotating negatively charged dust grain), in central gravitational and rotating dipole magnetic fields, corresponding in the zero approximation to the fields of magnetized compact star with $M=1.5M_{\odot}$, $R=10^6$cm, and {\bf (a)} $\Omega=0$, $B_0=10^3$G; {\bf (b)} $\Omega=10^3$rad/s, $B_0=10^8$G.
Positions of stable halo orbits correspond to the potential minima. Corresponding, stable in radial direction, circular equatorial orbit corresponds to the saddle point in the equatorial plane.} 
\end{figure}

\begin{figure}[t!]
\begin{center}
\includegraphics[width=0.75\hsize]{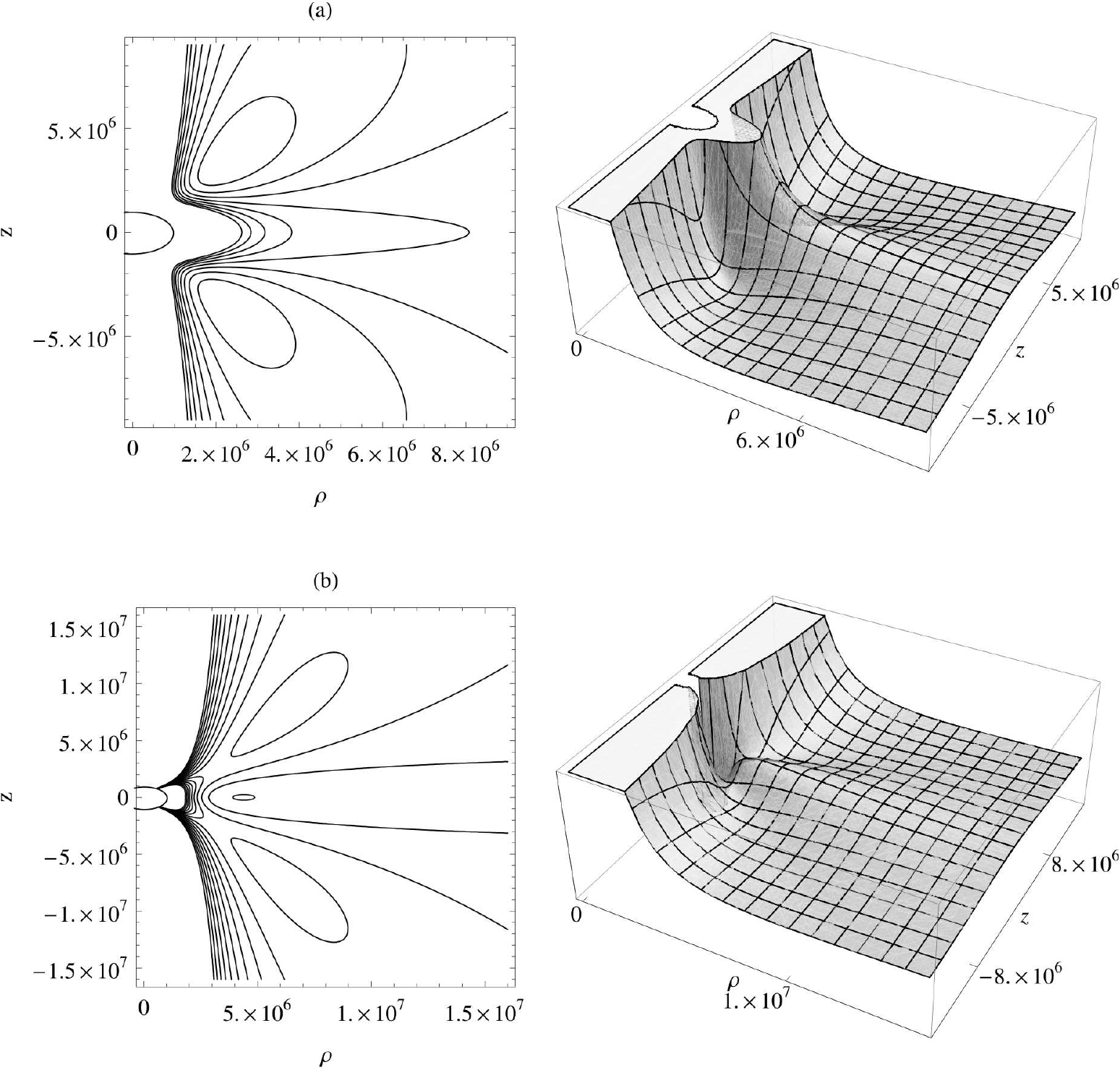}
\end{center}
\caption{\label{Fig:2} Effective potential $U_{\rm eff}$ and its contours for motion of particle with (in Gaussian units) {\bf (a)} $m=4.2\times10^{-12}$, $q=3.3\times10^{-5}$, $\bar{p}_{\phi}=-5\times10^{4}$ (counter-rotating positively charged dust grain) {\bf (b)} $m=4.2\times10^{-12}$, $q=3.3\times10^{-5}$, $\bar{p}_{\phi}=10\times10^{4}$ (co-rotating positively charged dust grain), in central gravitational and rotating dipole magnetic fields, corresponding in the zero approximation to the fields of magnetized compact star with $M=1.5M_{\odot}$, $R=10^6$cm and  $\Omega=7\times10^3$rad/s, $B_0=10^8$G.
Positions of stable halo orbits correspond to the potential minima. Concomitant unstable circular orbit corresponds to the local maximum in the case (b).} 
\end{figure}

The particle motion is described by Hamiltonian  
\begin{eqnarray}
\mathcal{H}=\frac{1}{2m}\left(\vec{p}-\frac{q}{c}\vec{A}\right)^2+\mathcal{U},
\end{eqnarray}
where $\mathcal{U}$ is the potential energy, and $p$, $m$, $q$ are, respectively, the momentum, mass and charge of the particle. 
In the cylindrical coordinates $(\rho,\phi,z)$, where $\rho=(x^2+y^2)^{1/2}$ and $\phi=\arctan(y/x)$,  
\begin{eqnarray}
\label{1}
\mathcal{H}=\frac{1}{2m}(p_{\rho}^2+p_z^2)+U_{\rm eff},
\end{eqnarray} 
with the effective potential   
\begin{eqnarray}
\label{2}
U_{\rm eff}=\frac{1}{2m\rho^2}(\bar{p}_{\phi}-\gamma\Psi)^2+\sigma_{\rm g}m\psi+\sigma_{\rm r}\gamma\Omega\Psi.
\end{eqnarray} 
In the above equation, $\bar{p}_{\phi}=m\rho^2d\phi/dt+\gamma\Psi$ is the conserved angular momentum, $\gamma=q\mathcal{M}/c$, and $\Psi=\rho^2/r^3$ is the dipole stream function. This function enables us to express the electric field in the form 
\begin{eqnarray}
\vec{E}=-\frac{1}{c}\mathcal{M}\Omega\nabla\Psi.
\end{eqnarray}
Following \cite{Dull-Hor-How:2002}, we also introduce \textquoteleft field switches\textquoteright~$\sigma_{\rm g}$ and $\sigma_{\rm r}$, taking values $0$ or $1$. 
Characteristic behaviour of the effective potential (\ref{2}) is shown in figures~\ref{Fig:1} and \ref{Fig:2} and suggests the existence of halo orbits near magnetized compact stars. 

Measuring time by $2\pi/\Omega$, distance by $\mathcal{R}=(GM/\Omega^2)^{1/3}$, and scaling the mass by the particle mass $m$, we find  
\begin{eqnarray}
\hat{H}&=&\frac{1}{2}(\hat{p}_{\rho}^2+\hat{p}_z^2)+\hat{U}_{\rm eff},\\
\hat{U}_{\rm eff}&=&\frac{1}{2}\left(\frac{\hat{p}}{\hat{\rho}}-\delta\frac{\hat{\rho}}{\hat{r}^3}\right)^2-\sigma_{\rm g}\frac{1}{\hat{r}-\hat{r}_{\rm S}}+\sigma_{\rm r}\delta\frac{\hat{\rho}^2}{\hat{r}^3},
\end{eqnarray} 
where the hat denotes scaled quantities, and  
\begin{eqnarray}  
\hat{p}=\bar{p}_{\phi}\frac{\mathcal{R}\delta}{\gamma},\quad \hat{r}_{\rm S}=\frac{2GM}{c^2\mathcal{R}},\quad\delta=\frac{\Omega\gamma}{mGM}.
\end{eqnarray}
Introducing the scaled spherical coordinates $(\hat{r}, \phi, \theta)$, where $\hat{r}=(\hat{\rho}^2+\hat{z}^2)^{1/2}$ and $\theta=\arccos(\hat{z}/\hat{r})$, and the orbital angular velocity of a particle
\begin{eqnarray}
\hat{\omega}=\frac{d\phi}{d\hat{t}}=\partial_{\hat{p}}\hat{U}_{\rm eff}=\frac{\hat{p}}{\hat{r}^2\sin^2{\theta}}-\frac{\delta}{\hat{r}^3},
\end{eqnarray}
we can write the effective potential in the form
\begin{eqnarray}
\label{18}
\hat{U}_{\rm eff}=\frac{1}{2}\hat{\omega}^2\hat{r}^2\sin^2{\theta}-\frac{\sigma_{\rm g}}{\hat{r}-\hat{r}_{\rm S}}+\frac{\sigma_{\rm r}\delta\sin^2{\theta}}{\hat{r}}.
\end{eqnarray}

Conditions for the existence of the halo orbits, $\partial_{\hat{r}} \hat{U}_{\rm eff}=0$ and $\partial_{\theta}\hat{U}_{\rm eff}=0$, are:
\begin{eqnarray}
-\hat{\omega}^2\hat{r}\sin^2{\theta}+\frac{\delta(\hat{\omega}-\sigma_{\rm r})\sin^2{\theta}}{\hat{r}^2}+\frac{\sigma_{\rm g}}{(\hat{r}-\hat{r}_{\rm S})^2}=0,\\
-\frac{1}{\hat{r}}\cos{\theta}\sin{\theta}(\hat{\omega}^2\hat{r}^3+2\hat{\omega}\delta-2\sigma_{\rm r}\delta)=0.
\end{eqnarray}
Eliminating $\hat{\omega}$ from these equations, we can write the halo orbits existence condition in the form   
\begin{eqnarray}
\label{20}
\delta=\frac{\pm\sigma_{\rm g}^2\hat{r}^3\csc{\theta}}{\sigma_{\rm g}(\hat{r}-\hat{r}_{\rm S})[(6\sigma_{\rm g}\hat{r})^{1/2}+3\sigma_{\rm r} \hat{r}(\hat{r}-\hat{r}_{\rm S})\sin{\theta}]}.
\end{eqnarray}
Thus, for arbitrarily fixed coordinates of halo orbits, there are real values of the parameter $\delta$, which can be set by tuning the object and particle parameters. 
From equation (\ref{20}), it is clear that the co-rotational electric field is not necessary for the existence of the halo orbits, i.e., the rotational effect of the magnetic field is not essential, although it influences the properties of the halo orbits. On the other hand, there are no halo orbits without the gravitational field. These conclusions are qualitatively the same as in  \cite{Dull-Hor-How:2002}. 

In the presence of the gravitational field and absence of the electric field, there are co-rotating stable halo orbits for particles with negative charge and counter-rotating orbits for particles with positive charge. The effective potential is invariant under simultaneous charge reversal and angular momentum reversal. In the case of presence of both the electric and magnetic fields, i.e., adding the rotation to the magnetic field, there are no qualitative changes for negatively charged particles motion along (always stable) halo orbits. Such particles must co-rotate with the source. On the other hand, the positively charged particles can either counter-rotate with the source or co-rotate, moving along stable halo orbits.\footnote{Preliminary   analysis of differences between the original Newtonian (weak gravity of a planet) and pseudo-Newtonian (strong gravity of a compact star) was presented in \cite{Sch-etal:2005}.}

\section{\label{sec:Schwarzschild}Schwarzschild black hole with dipole magnetic field}
We consider a Schwarzschild black hole of mass $M$ described by the metric \cite{Mis-Tho-Whe:1973}
\begin{eqnarray}
ds^2&=&-\left(1-\frac{2M}{r}\right)dt^2+\left(1-\frac{2M}{r}\right)^{-1}dr^2+r^2(d\theta^2+\sin^2{\theta}d\phi^2),
\end{eqnarray}
in the geometric units $(c=G=1)$ and standard Schwarzschild coordinates $(t,r,\phi,\theta)$. 
The black hole is surrounded, in the equatorial plane, by a toroidal current loop of radius $R$, which is supposed to contribute to the electromagnetic field but not to the gravitational field. The magnetic field generated by the loop is influenced by the black hole gravitation \cite{Pet:1974}. It can be properly described by the leading (dipole) term of the full multipole solution \cite{Pre:2004}, i.e., by  
\begin{eqnarray}
A_i=-\frac{3}{8}\delta^{\phi}_i\frac{\mathcal{M}r^2\sin^2{\theta}}{M^3}\left[\ln{\left(1-\frac{2M}{r}\right)}+\frac{2M}{r}+\frac{2M^2}{r^2}\right]
\end{eqnarray}
in the region $r>R$. Here $\mathcal{M}=\pi R^2I(1-2M/r)^{1/2}$
is magnitude of the dipole moment of the magnetic field and $I$ is the current in the loop.

\begin{figure}
\begin{center}
\includegraphics[width=0.75\hsize]{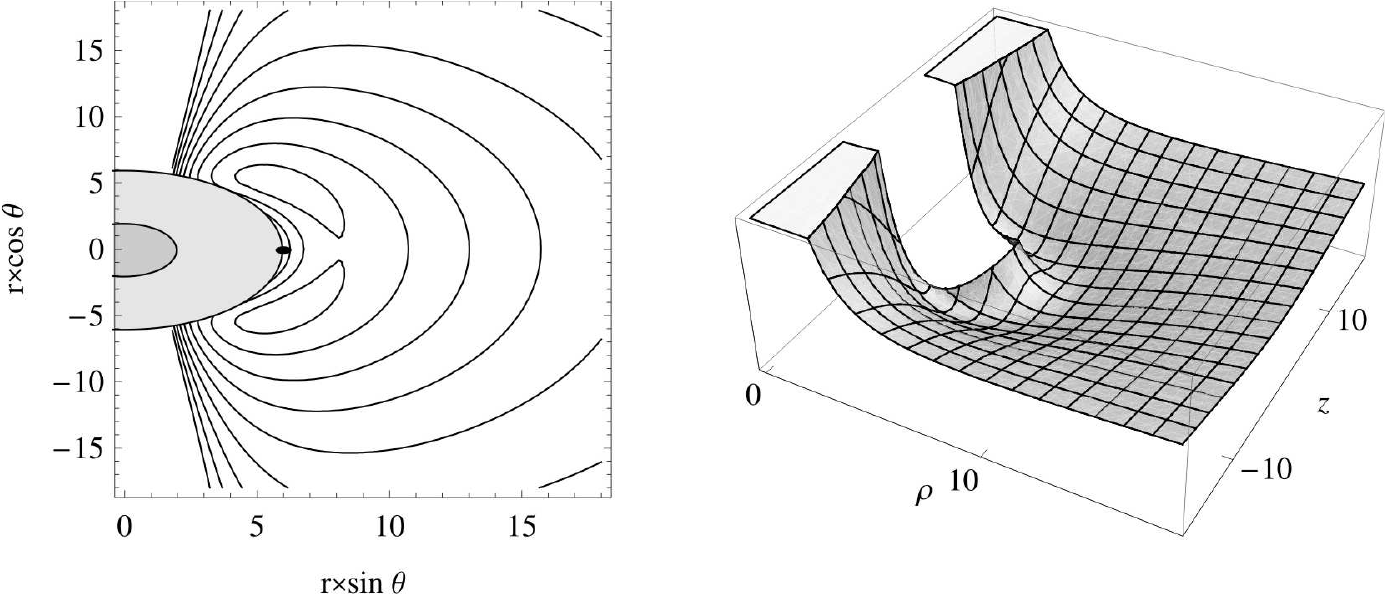}
\end{center}
\caption{\label{Fig:3} Effective potential $V_{\rm eff}$ and its contours for motion of charged particle with $\tilde{L}=1$ (in geometric units and units of $M$) in the Schwarzschild spacetime with the test dipole magnetic field of current loop characterized by the parameter $k=-4$.
Positions of stable halo orbits correspond to the potential minima. Concomitant equatorial circular orbit, stable in radial direction, corresponds to the saddle point in the equatorial plane. The effective potential is not relevant in the region limited from above by the radius of the current loop $R=6$ (light gray), entirely containing the dynamic region $r<2$ of the spacetime (dark gray).} 
\end{figure}

The \textquoteleft super Hamiltonian\textquoteright~for the motion of a test particle reads   
\begin{eqnarray}
\label{25}
\mathcal{H}=\frac{1}{2}g^{ij}(\bar{p}_i-qA_i)(\bar{p}_j-qA_j), 
\end{eqnarray} 
where $\bar{p}_i$ is the generalized momentum. The first Hamilton's equation implies $dx^i/d\lambda=\bar{p}^i-qA^i\equiv p^i$. The second Hamilton's equation ensures that the momenta
\begin{eqnarray}
\label{26}
\bar{p}_t&=&p_t+qA_t=-m\left(1-\frac{2M}{r}\right)\frac{dt}{d\tau}\equiv-E\\ 
\label{27}
\bar{p}_{\phi}&=&p_{\phi}+qA_{\phi}=mr^2\sin^2{\theta}\frac{d\phi}{d\tau}+qA_{\phi}\equiv L
\end{eqnarray}
are conserved.
Introducing now the specific energy, angular momentum and charge by the relations $\tilde{E}=E/m$, $\tilde{L}=L/m$ and $\tilde{q}=q/m$, we find
\begin{eqnarray}
\tilde{E}^2=({p^r})^2+r^2\left(1-\frac{2M}{r}\right)(p^{\theta})^2+V_{\rm eff}^2.
\end{eqnarray} 
The quantity $V_{\rm eff}$ is the effective potential taking in the dimensionless coordinates $r/M\rightarrow r$, $t/M\rightarrow t$ the form
\begin{eqnarray}
\label{300}
V^2_{\rm eff}&=&(1-\frac{2}{r})\left\{1+\left(\frac{\tilde{L}}{r\sin{\theta}}+k r\sin{\theta}\left[\ln{\left(1-\frac{2}{r}\right)}+\frac{2}{r}+\frac{2}{r^2}\right]\right)^2\right\},
\end{eqnarray}
where 
\begin{eqnarray}
\label{301}
k=\frac{3}{8}\tilde{q}\pi R^2I\left(1-\frac{2M}{r}\right)^{1/2}=const.
\end{eqnarray}
Figure~\ref{Fig:3} shows the characteristic behaviour of the effective potential (\ref{300}), suggesting the existence of halo orbits.  

In order to confirm the presence of halo orbits, one needs to check conditions for the existence of halo orbits $\partial_rV_{\rm eff}=0$ and $\partial_{\theta}V_{\rm eff}=0$, which now imply the \textquoteleft master\textquoteright~relation for the parameter $k$ 
\begin{eqnarray}
\label{30a}
k=\frac{\pm\frac{1}{2}r\csc{\theta}}{\sqrt{-[2+2r+r^2\ln{(1-\frac{2}{r})}]
[6+2r+r^2\ln{(1-\frac{2}{r})}]}}.
\end{eqnarray}
The reality condition in this relation restricts the halo orbits existence to the region $r>2.3231$. But this condition is automatically satisfied due to the assumption of the loop radius $R\geq6$ \cite{Pre:2004}. Thus, considering a circular orbit with the radius $r>R$ and latitude $0<\theta<\pi$, we can find corresponding value of the parameter $k$ and fit this value by tuning parameters of the particle, black-hole and magnetic field in equation (\ref{301}). The Schwarzschild black hole with a test dipole magnetic field and a non-rotating compact star with aligned (static) dipole magnetic field can be considered as equivalent to each other. As intuitively expected, there is the same type of behaviour of the relativistic effective potential (\ref{300}) and the pseudo-Newtonian one (\ref{2}) without the contribution of the electric field. This consistency in type of the effective potentials behaviour confirms our results from section~\ref{sec:CompactStar}. The other types of the effective potential (\ref{2}) behaviour are not recovered in the Schwarzschild case, simply because there are no rotational effects inducing the electric counterpart.

\section{\label{sec:Kerr-Newman}Kerr-Newman spacetime}
Now, we study halo orbits existence within Kerr-Newman solution of the full Einstein-Maxwell equations. 
In the dimensionless Boyer-Linquist coordinates $(t,r,\phi,\theta)$, the line element takes the form \cite{Mis-Tho-Whe:1973}
\begin{eqnarray}
\label{metKN}
ds^2&=&-\frac{\Delta}{\Sigma}(dt-a\sin{\theta}d\phi)^2+\frac{\Sigma}{\Delta}dr^2+\Sigma d\theta^2\nonumber\\&&+\frac{\sin^2{\theta}}{\Sigma}[(r^2+a^2)d\phi-adt]^2,
\end{eqnarray}  
where $\Delta=r^2-2r+a^2+e^2$ and $\Sigma=r^2+a^2\sin^2{\theta}$. Quantities $a$ and $e$ are the rotational and charge parameters of the spacetimes.  
Non-zero components of the antisymmetric electromagnetic field tensor $F_{ij}=A_{j,i}-A_{i,j}$ read
\begin{eqnarray}
\label{elamg1}
F_{rt}&=&\frac{e(r^2-a^2\cos^2{\theta})}{\Sigma^2},\\
\label{elmag2}
F_{r\phi}&=&\frac{-ae\sin^2{\theta}(r^2-a^2\cos^2{\theta})}{\Sigma^2},\\
\label{elmag3}
F_{\theta t}&=&\frac{-a^2er\sin{2\theta}}{\Sigma^2},\\
\label{elmag4}
F_{\theta\phi}&=&\frac{aer\sin{2\theta}(r^2+a^2)}{\Sigma^2}.
\end{eqnarray}

\begin{figure}[t!]
\begin{center}
\includegraphics[width=0.75\hsize]{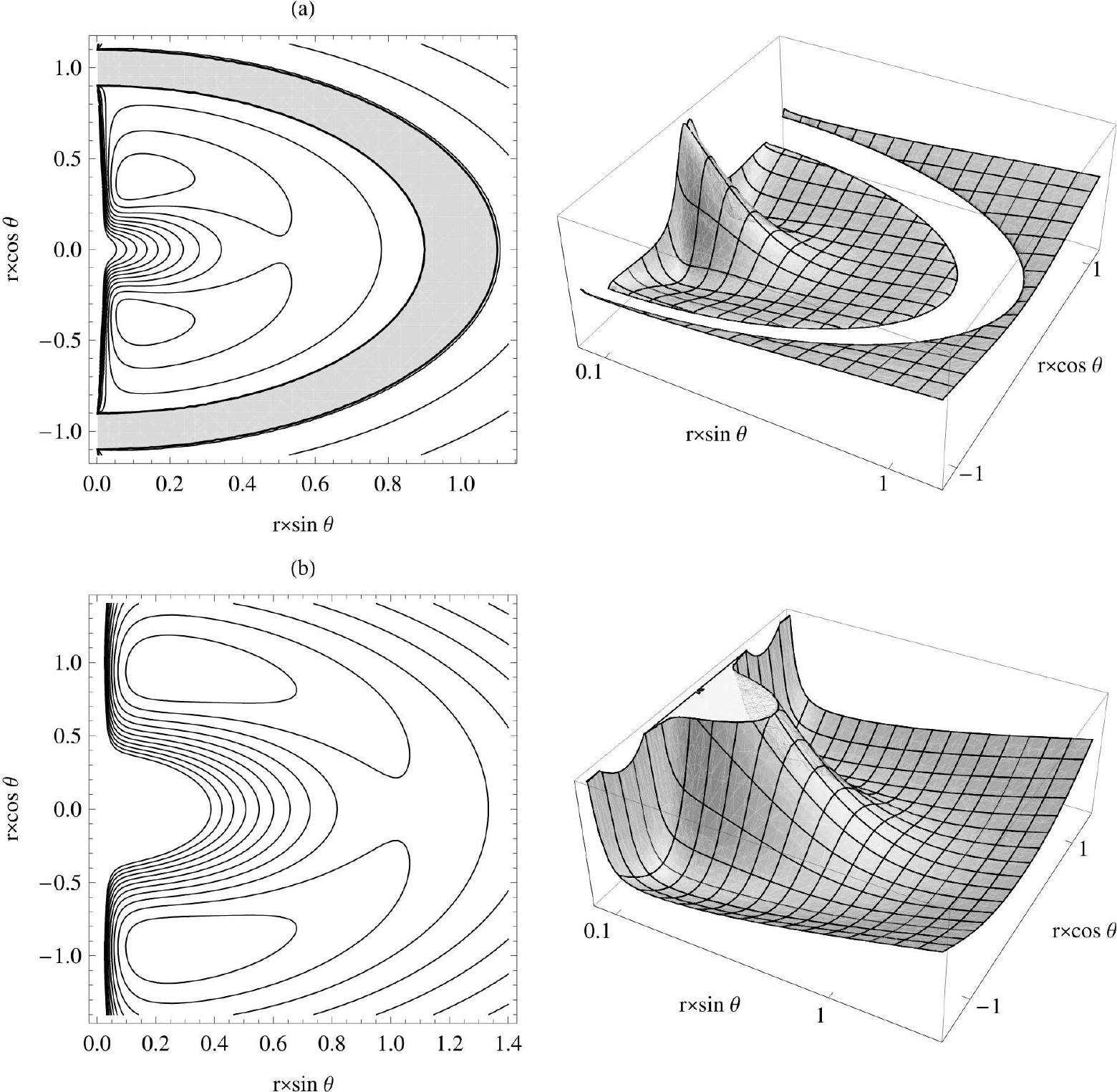}
\end{center}
\caption{\label{Fig:5} Effective potential $W_{\rm eff}$ and its contours for motion of charged particles with $\tilde{L}=1$ and $\tilde{q}=-5$ (in geometric units and units of $M$) in the Kerr-Newman {\bf (a)} black-hole spacetime with $a^2=0.04$ and $e^2=0.95$; {\bf (b)} naked-singularity spacetime with $a^2=0.6$ and $e^2=0.95$. 
Positions of stable halo orbits correspond to the potential minima. Concomitant equatorial circular orbit, stable in radial direction, corresponds to the saddle point in the equatorial plane. The effective potential $W_{\rm eff}$ is not relevant in the region between the event horizons (gray), where $\Delta<0$.} 
\end{figure}

\begin{figure}[t!]
\begin{center}
\includegraphics[width=0.75\hsize]{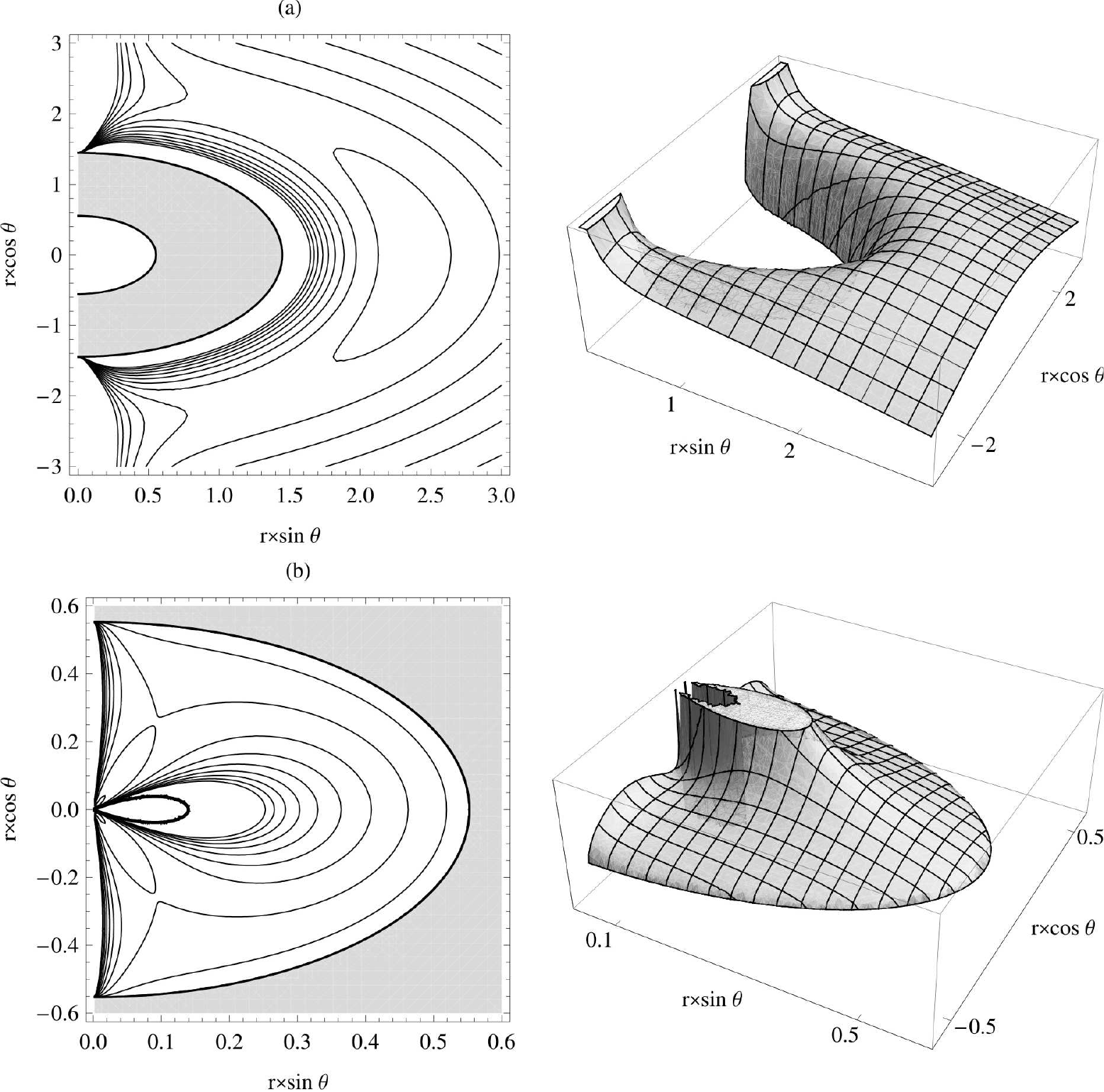}
\end{center}
\caption{\label{Fig:6} Effective potential $W_{\rm eff}$ and its contours for motion of charged particles with $\tilde{L}=0.1$ and $\tilde{q}=2.9$ (in geometric units and units of $M$) in the Kerr-Newman black-hole spacetime with $a^2=0.5$ and $e^2=0.3$, in the regions {\bf (a)} above the outer event horizon; {\bf (b)} under the inner event horizon. Positions of unstable stable halo orbits correspond to the saddle points of the potential. Concomitant equatorial unstable circular orbit corresponds to the potential maximum in the equatorial plane. 
The effective potential $W_{\rm eff}$ is not relevant in the region between the event horizons (gray), where $\Delta<0$.}
\end{figure}

\begin{figure}[t!]
\begin{center}
\includegraphics[width=0.75\hsize]{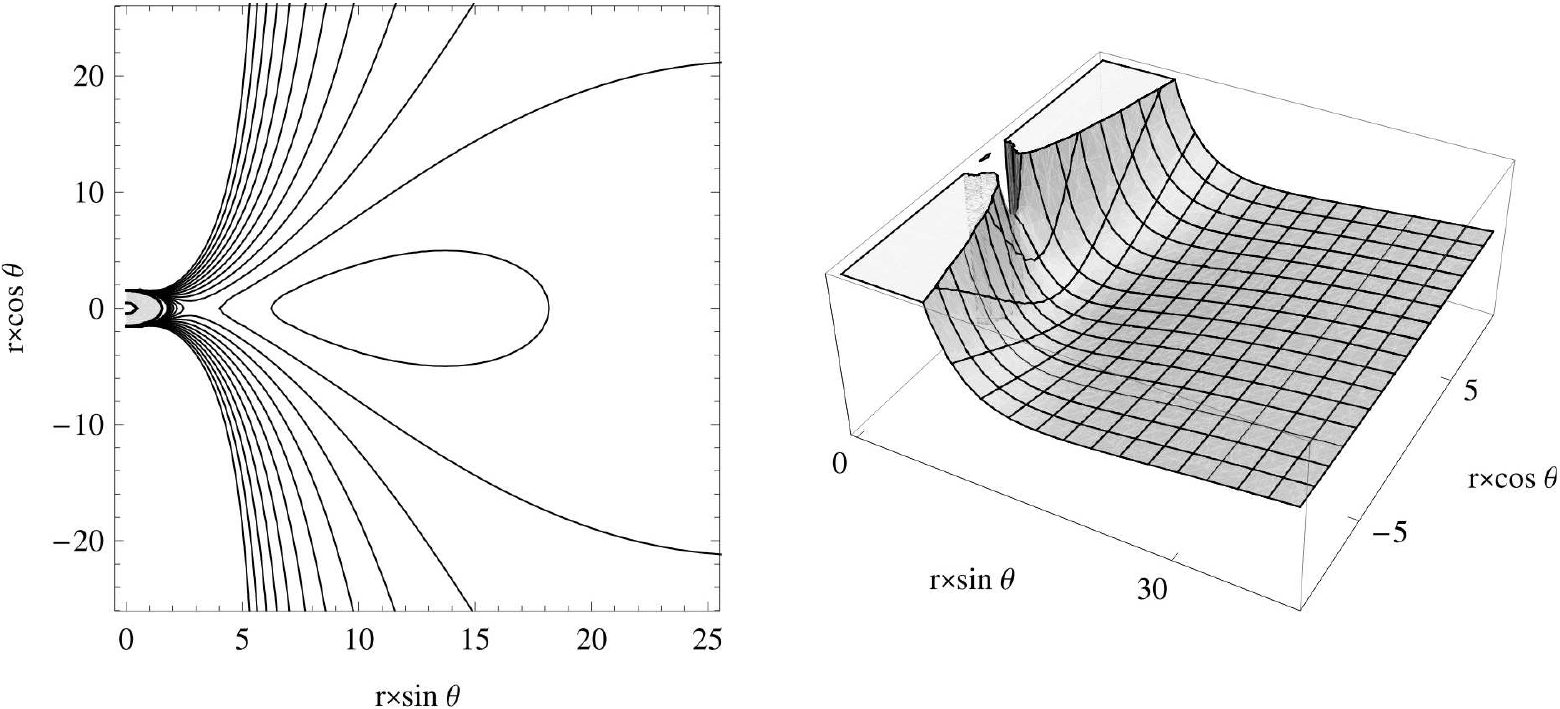}
\end{center}
\caption{\label{Fig:7} Effective potential $W_{\rm eff}$ and its contours for motion of charged particles with $\tilde{L}=8$ and $\tilde{q}=-5$ (in geometric units and units of $M$) in the region above the outer event horizon of the Kerr-Newman black-hole spacetime with $a^2=0.2$ and $e^2=0.5$. The potential minimum corresponds to the stable equatorial circular orbit. 
The effective potential $W_{\rm eff}$ is not relevant in the region between the event horizons (gray), where \mbox{$\Delta<0$}. 
In all numerically tested cases, no stable halo orbits have been found, only the stable equatorial circular orbits were confirmed.} 
\end{figure}

The effective potential for the motion of a particle with the specific charge $\tilde{q}=q/m$,  takes the form \cite{Mis-Tho-Whe:1973,Car:1968}
\begin{eqnarray}
\label{potential}
W_{\rm eff}=\frac{Y+\sqrt{Y^2-XZ}}{X},
\end{eqnarray}
where 
\begin{eqnarray}
X&=&(r^2+a^2)^2-\Delta a^2\sin^2{\theta},\\
Y&=&(\tilde{L}a+\tilde{q}er)(r^2+a^2)-\tilde{L}a\Delta,\\
Z&=&(\tilde{L}a+\tilde{q}er)^2-\Delta\Sigma-\Delta\tilde{L}^2/\sin^2{\theta},
\end{eqnarray}
and $\tilde{L}=L/m$ is the conserved axial component of specific angular momentum. 
Characteristic behaviour of the effective potential (\ref{potential}) suggesting the existence of halo and equatorial orbits in Kerr-Newman spacetimes is shown in figures~\ref{Fig:5}--\ref{Fig:7}.   

In order to study the halo orbits, we find it very helpful to employ the \textquoteleft force approach\textquoteright~\cite{Abr-Car-Las:1988}.\footnote{We found \cite{Kov-Stu:2007} the force analysis of circular geodesics to be much more efficient and straightforward in comparison with the effective potential approach.}  
For uniform circular motion of a particle at constant radius and latitude, and with the velocity $v$ measured by ZAMO \cite{Bar-Pre-Teu:1972}, we can write (see Appendix A) two force equations: 
\begin{eqnarray}
\label{40}
-\mathcal{G}_{r}-(\gamma v)^2\mathcal{Z}_{r}-\gamma^2v\mathcal{C}_{r}=\tilde{q}\gamma(\mathcal{E}_{r}+v\mathcal{M}_{r}),\\
\label{41}
-\mathcal{G}_{\theta}-(\gamma v)^2\mathcal{Z}_{\theta}-\gamma^2v\mathcal{C}_{\theta}=\tilde{q}\gamma(\mathcal{E}_{\theta}+v\mathcal{M}_{\theta}),
\end{eqnarray}
where Lorentz factor $\gamma=(1-v^2)^{-1/2}$.
Here $\mathcal{G}_j$, $\mathcal{Z}_j$, $\mathcal{C}_j$ denote the mass and velocity independent parts of the gravitational, centrifugal and Coriolis inertial forces, and $\mathcal{E}_j$ and $\mathcal{M}_j$ stand for the charge and velocity independent parts of the electric and magnetic forces (see relations (\ref{A15})--(\ref{A24})).

\begin{figure}
\begin{center}
\includegraphics[width=0.75\hsize]{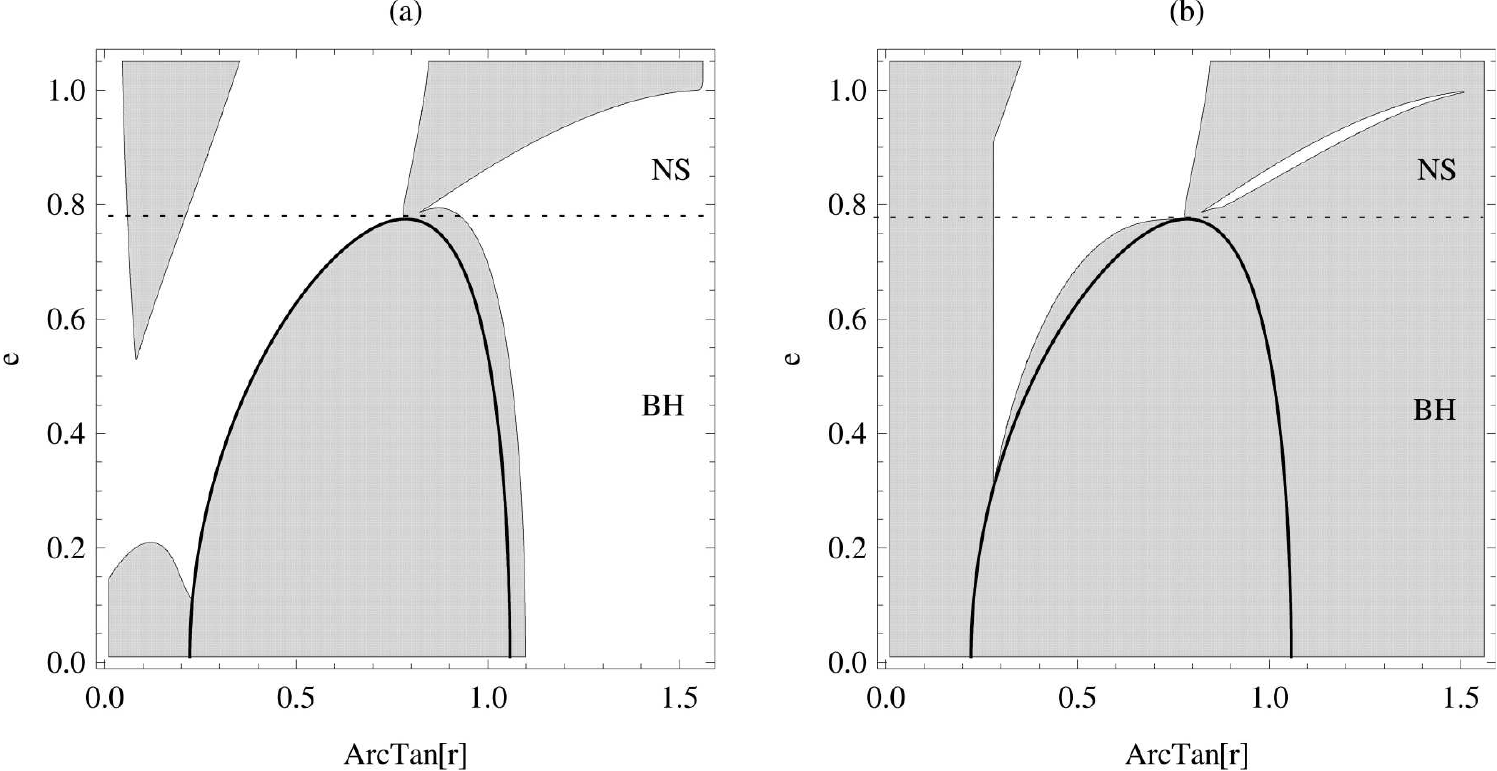}
\end{center}
\caption{\label{Fig:4} Existence of halo orbits for $\theta=1.1$ in Kerr-Newman black-hole (BH) and naked-singularity (NS) spacetimes with $a^2=0.4$. Positions of horizons are denoted by the thick curve. {\bf (a)} White areas correspond to the regions where the root $v_I$ is real and $-1<v_I<1$, i.e., where the halo orbits occur. {\bf (b)} White areas correspond to the regions where the stationary points of the effective potential $W_{\rm eff}$, corresponding to the values of $v_I$, are minima, i.e., where the halo orbits are stable.} 
\end{figure}

Eliminating $\tilde{q}$ from equations (\ref{40}) and (\ref{41}), and assuming $0<\theta<\pi/2$, i.e., omitting stationary equilibrium points on the axis of symmetry and circular orbits in the equatorial plane, we obtain a cubic equation 
\begin{eqnarray}
\label{50}
Av^3+Bv^2+Cv+D=0,
\end{eqnarray}
where
\begin{eqnarray}
A&=&\mathcal{M}_{\theta}(\mathcal{G}_r-\mathcal{Z}_r)+\mathcal{M}_{r}(\mathcal{Z}_{\theta}-\mathcal{G}_{\theta}),\\
B&=&\mathcal{E}_{r}(\mathcal{G}_{r}-\mathcal{Z}_{r})+\mathcal{E}_{r}(\mathcal{Z}_{\theta}-\mathcal{G}_{\theta})
+\mathcal{C}_{\theta}\mathcal{M}_{r}-\mathcal{C}_{r}\mathcal{M}_{\theta},\\
C&=&\mathcal{C}_{\theta}\mathcal{E}_{r}-\mathcal{C}_{r}\mathcal{E}_{\theta}+\mathcal{G}_{\theta}\mathcal{M}_{r}-\mathcal{G}_{r}\mathcal{M}_{\theta},\\
D&=&\mathcal{E}_{r}\mathcal{G}_{\theta}-\mathcal{E}_{\theta}\mathcal{G}_{r}.
\end{eqnarray}
This gives three, in general complex, solutions $v_I(r,\theta;a,e)$, $v_{II}(r,\theta;a,e)$ and $v_{III}(r,\theta;a,e)$ for possible orbital velocities of charged particles moving along the halo orbits.  
Parameterizing them by the spin $a$ and latitude $\theta$, we can investigate the validity of the condition $v_i\in R$ and $-1<v_i<1$ $ (i=I,II,III)$, in the plane $(r\times e)$. Our numerical analysis of $v_i$ confirms the existence of halo orbits in the Kerr-Newman naked-singularity spacetime as well as in the black-hole spacetimes under the inner event horizon and even over the outer horizon (see the representative case in figure~\ref{Fig:4}). 

We conclude our study of halo orbits existence by the discussion of their stability, namely by seeking for the stable ones. In the previous two cases (sections~\ref{sec:CompactStar} and \ref{sec:Schwarzschild}), the stability of the halo orbits could be directly inferred from the behaviour of related effective potentials. For the present purpose, we can use equation (\ref{40}) and express the specific charge $\tilde{q}_{\rm h}$ of the particle moving along the expected halo orbit at velocity $v$ by the relation
\begin{eqnarray}
\label{52}
\tilde{q}_{\rm h}=\frac{\mathcal{G}_r(v^2-1)-v(\mathcal{C}_r+v\mathcal{Z}_r)}{(\mathcal{E}_r+v\mathcal{M}_r)\sqrt{1-v^2}}.
\end{eqnarray}
Condition (\ref{52}) provides three possible values $\tilde{q}_{{\rm h},i}$ related to $v_i$.
By rewriting relation (\ref{27}) for the Kerr-Newman case and using the \mbox{4-velocity} decomposition (\ref{A1}), we can write for the specific angular momentum of particles at halo orbits
\begin{eqnarray}
\label{55}
\tilde{L}_{\rm h}=\gamma v \sqrt{g_{\phi\phi}}+\tilde{q}_{\rm h}A_{\phi},   
\end{eqnarray}
obtaining three possible values $\tilde{L}_{{\rm h},i}$ related to $\tilde{q}_{{\rm h},i}$ and $v_i$.

The stable halo orbits must satisfy the conditions of effective potential minima, i.e., 
\begin{eqnarray}
\label{60}
\partial^2_r W_{\rm eff}(r,\theta;a,e,\tilde{L}=\tilde{L}_{\rm h},\tilde{q}=\tilde{q}_{\rm h})>0,\\
\label{61}
\partial^2_{\theta} W_{\rm eff}(r,\theta;a,e,\tilde{L}=\tilde{L}_{\rm h},\tilde{q}=\tilde{q}_{\rm h})>0,
\end{eqnarray}
which we analyze in a numerical way in the plane $(r\times e)$, considering the parameterization by $a$ and $\theta$, for all three pairs of $\tilde{L}_{{\rm h},i}$ and $\tilde{q}_{{\rm h},i}$. 
Searching systematically through the parameter space of $(r\times e \times a \times \theta)$, we found no combinations of parameters allowing the existence of stable halo orbits in the Kerr-Newman black-hole spacetimes above the outer event horizon. Representative results of such analysis are shown in figure~\ref{Fig:4}, where both black holes and naked singularities are explored. In these plots we keep the spin $a$ fixed, whereas the electric charge $e$ varies along the ordinate. We find this type of plots very instructive, as they capture the entire range of radii from the origin up to spatial infinity. Shading distinguishes the areas where the halo orbits cannot exist (left panel), or where they are unstable (right panel). 
On the other hand, the region under the inner event horizon and the naked-singularity spacetimes do exhibit stable halo orbits. 

\section{\label{sec:Discussion}Discussion}
Our analysis has shown that the stable halo orbits can exist near rotating as well as non-rotating magnetized compact stars. Rotation enriches the classification of the stable halo orbits types. Our first example,  the pseudo-Newtonian gravitational potential and the classical dipole
magnetic field do not describe the situation with the same precision as a relevant solution of Einstein-Maxwell equations, especially in the region under the photon circular orbit, where the Paczynski-Wiita potential does not describe the gravitational field properly. However, the pseudo-Newtonian approach is sufficient to demonstrate how the halo orbits arise in strong gravity. The halo orbits can exist even far away from the photon circular orbit, for realistic values of the particle specific charge and the compact star magnetic field strength. Moreover, at such distances, the gravitational and magnetic fields of one of the investigated cases, i.e., the slowly rotating magnetized compact star, are nearly of the same character as the fields generated by Schwarzschild black hole with a current loop in the equatorial plane, our second example. The general relativistic study of such configuration has confirmed the existence of stable halo orbits. 

In the fields of Kerr-Newman black holes, the third example, no stable halo orbits have ever been, to our knowledge, found outside the outer horizon, although there is a magnetic field of dipole character as well. But here, in contrast to the case of magnetized compact stars and Schwarzschild black holes with the externally generated magnetic field, where the magnetic field strength is generally independent of the spacetime parameters, the magnetic (and related electric) field is determined by the internal spacetime parameters (the spin and electric charge) and cannot be imposed as an independent variable. It seems that the connection of the electromagnetic field with the spin of the spacetime prevents the existence of stable off-equatorial circular orbits outside the outer horizon. Only unstable halo orbits are allowed there. On the other hand, the combinations of spacetime parameters allow the existence of stable halo orbits under the inner horizon of black holes. The inner horizon is also a Cauchy horizon that may focus to a spacelike one \cite{Bur:1997}. This may imply that the continuation of the spacetime manifold beyond it lacks any physical or astronomical relevance. However, beyond the ring singularity of Kerr metric, there is an asymptotically flat universe. If we actually live in such an asymptotically flat region, in addition to other interesting phenomena like closed timelike curves, one may also observe the halo orbits. Therefore, the existence and properties of halo orbits may be a useful tool probing the internal properties of black holes along with the traditional analysis of their geodesic structure \cite{Poi-Isr:1990}.       
In principle, as follows from our analysis, halo orbits become accessible in the naked-singularity spacetimes as well. Therefore this fact may be important in the identification of naked singularities. 

These conclusions are consistent with the results of a purely analytical study of the situation on the axis of symmetry, where the circular motion degenerates to a point. There, only unstable positions are allowed above the outer event horizon. The stable positions occur only under the inner horizon or in the naked-singularity spacetimes \cite{Bic-Stu-Bal:1989}. 
Thus, within the framework of the Kerr-Newman black-hole spacetimes, the only astrophysically relevant stable circular orbits at constant radius and latitude are those in the equatorial plane
\cite{Dad-Kal:1977}.   

The equatorial plane is where a gaseous accretion disk can reside. Although the accretion discs are often described as geometrically thin and planar structures, in reality they must have extended atmospheres which reach far above and below the equatorial plane. This motivated us to speculate that the halo orbits could be relevant for accretion in strong gravity. 
High frequency kilohertz quasiperiodic oscillations (QPOs) observed in some microquasars and binary systems with compact stars are frequently explained by variety of models based on the equatorial quasicircular motion, with characteristic orbital Keplerian and epicyclic frequencies. Most promising seem to be the relativistic precession model \cite{Ste-Vie:1999}, and the orbital resonance model \cite{Klu-Abr:2001,Ali-Gal:1981} or its generalization to the orbital multiresonant model \cite{Stu-Kot-Tor:2007,Tor-Bak-Stu-Sra:2007} for both binary systems with black holes (microquasars) and compact stars; in the case of near-extreme Kerr black-hole candidates (e.g. well known GRS 1915+105 microquasar) the complex high frequency QPOs patterns could be explained using the extended resonance model with the so-called hump-induced oscillations, additional to the orbital epicyclic oscillations \cite{Stu-Sla-Tor:2007a,Stu-Sla-Tor:2007b}. By using the orbital resonance models, black hole parameters, especially the spin, could be determined \cite{Tor:2005a,Tor:2005b,Klu-Abr-Bur-Tor:2007}. 

The existence of two mutually detached off-equatorial lobes of bound stable orbits can have profound consequences for plasma oscillations near compact objects. It allows us to imagine a situation when the bulk motion is along the stable circular orbits in the center of the lobes, around which a fraction of trapped particles oscillate. We imagine that the oscillations modulate the electromagnetic  signal from plasma cloud, and this could be observed as a mode of oscillations in the detected radiation.  

The halo orbits in the magnetic field of compact stars and black holes could be related to the oscillatory motion with \textquoteleft halo\textquoteright~radial and vertical frequencies, which could be considered as complementary model of the QPOs observed in the binary system of compact stars and microquasars. 

\section{\label{sec:Conclusions}Conclusions}
Our study has shown that a strong gravitational field endowed with a magnetic field of dipole character enable the existence of the so-called stable halo orbits, i.e., stable circular off-equatorial orbits. Such a composition of fields is expected in the vicinity of compact objects, e.g., in physically relevant situations with magnetized compact (neutron, quark) stars, black holes with current loops in the equatorial plane, or in the case of charged and rotating black holes. 
We have found that the stable halo orbits can take place near all of the three mentioned kinds of compact objects. But in the Kerr-Newman charged and rotating black-hole spacetimes, the stable halo orbits are at most of marginal astrophysical importance, being hidden under the event horizon. We  mentioned the possibility that the observer could actually live in an asymptotically flat region beyond the ring singularity, however, this would be clearly a highly speculative proposal. On the other hand, in the field of magnetized neutron stars or near black holes with current loops in the equatorial plane, the stable halo orbits appear outside the body and can be astrophysically relevant.

\ack
This work was supported by the Czech grant MSM 4781305903 and AV0Z10030501. V. Karas and Z. Stuchl\'ik thank also the Czech Science Foundation (GA\v{C}R 205/07/0052 and GA\v{C}R 202/06/0041). The authors thank anonymous referees for their comments and suggestions.

\appendix
\section{Projection of Lorentz equation and inertial forces formalism}
We start with the decomposition of the particle \mbox{4-velocity} into the form
\cite{Abr-Nur-Wex:1995}
\begin{eqnarray}
\label{A1}
u^i=\gamma(n^i+v\tau^i),
\end{eqnarray}
where the \mbox{4-velocity} field $n^i$ satisfies the conditions 
\begin{eqnarray}
\label{A2}
n^k n_k=-1,\quad
n^i\nabla_i n_k=\nabla_k\Phi,\quad n_{[i}\nabla_j n_{k]}=0,
\end{eqnarray}
and the vector $\tau^i$ is the unit spacelike vector orthogonal to it, along which the spatial \mbox{3-velocity} with magnitude $v$ is aligned. The \mbox{4-velocity} field $n^i$ can be chosen in the form
\begin{eqnarray}
\label{A3}
n^i=e^{-\Phi}\iota^i,\quad e^{2\Phi}=-\iota^i\iota_i,
\end{eqnarray}
thus it corresponds to \mbox{4-velocity} of stationary observers, parallel to a timelike vector field $\iota^i$. 

Now, we can express the left-hand side of the Lorentz equation 
\begin{eqnarray}
\label{30}
mu^{k}\nabla_{k}u_{j}=qF_{jk}u^{k}
\end{eqnarray}
in the form 
\begin{eqnarray}
\label{A4}
ma_k&=&m[\gamma^2\nabla_k\Phi+\gamma^2 v(n^i\nabla_i\tau_k+\tau^i\nabla_{i}n_k)\nonumber\\
&&+\gamma^2v^2\tau^i\nabla_i\tau_k+(v\gamma\dot{)}\tau_k+\dot{\gamma}n_k],
\end{eqnarray}
where $(v\gamma\dot{)}=u^i\nabla_i(\gamma v)$. 
By using the projection tensor  
\begin{eqnarray}
\label{A5}
h_{ik}=g_{ik}+n_i n_k,
\end{eqnarray}
we project the uniquely decomposed \mbox{4-force} (\ref{A4}): 
\begin{eqnarray}
\label{A6}
ma_j^{\perp}=mh^k_ja_k=-{G_j}-{Z_j}-{C_j}-{L_j},
\end{eqnarray}
where 
\begin{eqnarray}
\label{A7}
G_j&=&-m\nabla_j\Phi,\\
\label{A8}
Z_j&=&-m(\gamma v)^2\tilde{\tau}^i\tilde{\nabla}_{i}\tilde{\tau}_j,\\
\label{A9}
C_j&=&-m\gamma^2vX_j,\\
\label{A10}
L_j&=&-m\dot{V}\tilde{\tau}_j
\end{eqnarray}
can be interpreted as the gravitational, centrifugal, Coriolis and Euler inertial forces \cite{Abr-Nur-Wex:1995}. 
In the above relations $X_j=n^i(\nabla_i\tau_j-\nabla_j\tau_{i})$ and $\dot{V}=-u^i\nabla_i(\iota^ku_kv)$.
The vector $\tilde{\tau}^{i}=e^{\Phi}\tau^{i}$, with its covariant form $\tilde{\tau}_i=e^{-\Phi}\tau_i$, is the spacelike unit vector parallel to $\tau^i$ in the so-called optical reference geometry, defined by the relation 
\begin{eqnarray}
\label{A11}
\tilde{h}_{ik}=e^{-2\Phi}h_{ik}.
\end{eqnarray}

Projecting the right-hand side of the Lorentz equation (\ref{30}), we obtain the equation \cite{Agu-etal:1996}
\begin{eqnarray}
\label{A11a}
q h^i_j F_{ik}u^k=E_j+M_j,
\end{eqnarray}
where 
\begin{eqnarray}
\label{A11b}
E_j&=&q\gamma F_{jk}n^j,\\
\label{A11c}
M_j&=&q\gamma v(F_{jk}\tau^k+n_jF_{kl}n^k\tau^l)
\end{eqnarray}
are the electric and magnetic forces, respectively.

We focus our attention to the case of the symmetric and stationary spacetimes with the Killing vector fields
$\eta^{i}=\delta^{i}_t$ and $\xi^{i}=\delta^{i}_{\phi}$, and consider $n^i$ to be \mbox{4-velocity} of the ZAMO observers, i.e.,  
\begin{eqnarray}
\label{A12}
n^i&=&e^{-\Phi}(\eta^i+\Omega_{\rm ZAMO}\xi^i),\\
e^{2\Phi}&=&-(\eta^i+\Omega_{\rm ZAMO}\xi^i)(\eta_i+\Omega_{\rm ZAMO}
\xi_i),
\end{eqnarray}
where $\Omega_{\rm ZAMO}=-\eta^{i}\xi_{i}/\xi^{j}\xi_{j}$.
Considering uniform circular motion of particle at constant radius and constant latitude, i.e.,
\begin{eqnarray}
\label{A14}
\tau^i=(\xi^k\xi_k)^{-1/2}\xi^i,
\end{eqnarray}
in Kerr-Newman spacetime, the mass and velocity independent parts of the non-zero radial and latitudinal components of the inertial forces (\ref{A7})--(\ref{A10}) take the form:
\begin{eqnarray}
\label{A15}
\mathcal{G}_r&=&-\frac{(a^2+r^2)[(r-1)(a^2+r^2)-2r\Delta]}{\Delta\nu}
-\frac{4r}{\mu}+\frac{r}{\Sigma},\\
\label{A16}
\mathcal{Z}_r&=&-\frac{2(a^2+r^2)[(r-1)(a^2+r^2)-2r\Delta]}{\Delta\nu}
-\frac{4r}{\mu}+\frac{r-1}{\Delta},\\
\label{A17}
\mathcal{C}_r&=&\frac{2a\sin{\theta}}{\mu\nu\sqrt{\Delta}}\{-r\Delta(3a^2+4r^2+a^2\cos{2\theta})\nonumber\\
&&+2(a^2+r^2)[(2r-1)(a^2+r^2)
-a^2(r-1)\sin^2{\theta}]\},
\end{eqnarray}
and
\begin{eqnarray}
\label{A18}
\mathcal{G}_{\theta}&=&\frac{1}{2}a^2\sin{2\theta}\left(\frac{4}{\mu}-\frac{\Delta}{\nu}-\frac{1}{\Sigma}\right),\\
\label{A19}
\mathcal{Z}_{\theta}&=&\cot{\theta}+a^2\sin{2\theta}\left(\frac{2}{\mu}-\frac{\Delta}{\nu}\right),\\
\label{A20}
\mathcal{C}_{\theta}&=&\frac{4a^3\cos{\theta}\sin^2{\theta}(\Delta-a^2-r^2)\sqrt{\Delta}}{\mu\nu},
\end{eqnarray}
where 
\begin{eqnarray}
\mu&=&a^2+2r^2+a^2\cos{2\theta},\\
\nu&=&(a^2+r^2)^2-\Delta a^2\sin^2{\theta}.
\end{eqnarray}
Euler force $L_k$ vanishes because of the uniformity of the motion.
The charge and velocity independent parts of the non-zero electric and magnetic force components (\ref{A11b}) and (\ref{A11c}) are given by the relations
\begin{eqnarray}
\label{A21} 
\mathcal{E}_{r}&=&\frac{e(a^2+r^2)(r^2-a^2\cos^2{\theta})}{\Sigma^{3/2}\sqrt{\nu\Delta}},\\
\label{A22}
\mathcal{M}_{r}&=&\frac{ae\sin{\theta}(a^2\cos^2{\theta}-r^2)}{\Sigma^{3/2}\sqrt{\nu}},\\
\label{A23}
\mathcal{E}_{\theta}&=&\frac{-2a^2er\sqrt{\Delta}\sin{2\theta}}{\mu\Sigma^{1/2}\sqrt{\nu}},\\
\label{A24}
\mathcal{M}_{\theta}&=&\frac{2aer\cos{\theta}(r^2+a^2)}{\Sigma^{3/2}\sqrt{\nu}}.
\end{eqnarray}

\end{document}